\documentclass[12pt,aps,prd,a4paper,superscriptaddress,nofootinbib,onecolumn]{revtex4-2}
\usepackage{amsfonts}
\usepackage{amsmath}
\usepackage{babel}
\usepackage[active]{srcltx}
\usepackage{graphicx}
\usepackage{color}
\usepackage{float}
\usepackage{changebar}
\usepackage{esint}
\usepackage{dsfont}
\usepackage{ae}
\usepackage{multirow}
\usepackage[usenames,dvipsnames]{xcolor}
\usepackage[colorlinks=true,pdfstartview=FitV,linkcolor=blue,citecolor=blue,urlcolor=blue,breaklinks=true]{hyperref}
\usepackage{braket}
\usepackage{mathtools}
\usepackage{slashed}
\usepackage{ulem}
\usepackage{empheq}
\usepackage{multirow}
\usepackage[utf8]{inputenc}
\usepackage[T1]{fontenc}
\usepackage{booktabs}

\newcommand{\beq}{\begin{equation}}
\newcommand{\eeq}{\end{equation}}
\newcommand{\beqa}{\begin{eqnarray}}
\newcommand{\eeqa}{\end{eqnarray}}

\begin{document}

\date{\today}

    \title{ Chirality loss during brane merging: a universal power law
       from the Jackiw-Rebbi index}

\author{H.~P.~Pinheiro}
\email{hudsonpinheiro@ufersa.edu}
\affiliation{Centro de Ci\^{e}ncias Exatas e Naturais, Universidade Federal Rural do Semi-\'{A}rido, Mossor\'{o}-RN, 59625-900, Brazil}
\author{C.~A.~S.~Almeida}
\affiliation{Departamento de F\'isica, Universidade Federal do Cear\'a (UFC),
Campus do Pici, Fortaleza-CE, 60455-760, Brazil}
\email{carlos@fisica.ufc.br}


\begin{abstract}
We investigate the rate at which chiral fermion localisation is lost
when two domain walls merge in extra-dimensional braneworld scenarios,
using the $(1+1)$-dimensional Jackiw-Rebbi framework as a controlled
analytical laboratory.
As the inter-brane separation $d$ decreases, left- and right-handed
zero modes hybridise and chiral asymmetry is progressively lost.
We show that the spatial separation between the chiral zero modes
follows a universal power law $|\Delta_{\mathrm{abs}}|\propto d^{\gamma}$
in the merging limit $d\to 0^{+}$, with the critical exponent $\gamma$
determined solely by the Jackiw-Rebbi topological index $N_{\mathrm{JR}}$,
and independent of the fermionic mass gap, the integrability of the
scalar sector, and the detailed shape of the domain wall profile.
Comparing the integrable sine-Gordon model with four members of the
non-integrable double sine-Gordon family, all sharing $N_{\mathrm{JR}}=1$,
we find $\gamma\in[0.930,0.985]$.
For the sine-Gordon model we derive the closed-form overlap integral
$I(d)=2d/\sinh(2d)$, from which the exact chiral separation follows as a ratio of hyperbolic functions without free parameters. This result identifies $\gamma$ as the crossover plateau of a local effective exponent $\gamma_{\mathrm{eff}}(d)$, explaining the sub-unit value analytically and tracing the universality to the
P\"{o}schl-Teller structure of the $N_{\mathrm{JR}}=1$ zero mode.
The universality of $\gamma$ implies that the rate of four-dimensional
Yukawa coupling collapse during brane merging is a topological
invariant, insensitive to the microscopic scalar dynamics generating
the walls.
\end{abstract}

\maketitle



\section{Introduction}
\label{sec:intro}
The localization of matter fields on topological defects is one of the central
mechanisms of extra-dimensional physics. The original proposal of Rubakov and
Shaposhnikov~\cite{Rubakov:1983bb} and Akama~\cite{Akama:1982} established
that Standard Model fields can be trapped on a four-dimensional brane embedded
in a higher-dimensional bulk, with the trapping achieved entirely by the
topology of the scalar field background rather than by explicit boundary
conditions. The fermionic sector of this programme rests on the seminal result
of Jackiw and Rebbi~\cite{Jackiw:1976}, who showed that a Dirac fermion
coupled to a kink background in $(1+1)$ dimensions develops a topologically
protected zero mode localised at the domain wall. Rubakov and Shaposhnikov
elevated this mechanism to $(4+1)$ dimensions: left- and right-handed fermions
couple with opposite chiralities to the wall, and the overlap of their
wave-functions in the extra dimension controls the effective four-dimensional
Yukawa couplings and, consequently, the fermion mass
hierarchy~\cite{ArkaniHamed:2000,Grossman:1999}.

A central and largely open question in this programme is how the localisation
properties of the chiral zero modes depend on the internal structure of the
brane. Realistic brane models are rarely described by a single thin domain
wall: the deformation method of Bazeia \textit{et al.}~\cite{Bazeia:2002}
generates smooth, thick-brane solutions with multikink structure, including
configurations that interpolate continuously between a single-brane and a
two-brane system. In this multi-wall setting, two physically distinct
parameters become relevant: the asymmetry of the scalar background and the
separation between constituent domain walls. Each parameter controls a
different aspect of fermion localisation, and their interplay determines
whether a chiral fermion spectrum can emerge from the extra dimension.

The interaction between fermionic zero modes and multi-kink backgrounds has
been studied in the context of bound-state spectra~\cite{Bazeia:2017} and
gravity localisation in generalised braneworld
scenarios~\cite{Bajc:2000,Bazeia:2009}. In particular, the hybridisation of
individual zero modes as domain walls merge — the limit in which the inter-wall
separation vanishes — is directly relevant to moduli stabilisation in two-brane
models~\cite{DeWolfe:2000,Gremm:2000}: if the inter-brane distance is a
dynamical modulus, the chiral structure of the four-dimensional spectrum
changes as the modulus evolves, and a quantitative description of this change
is needed for any phenomenological analysis.

Recently, a power-law scaling $|\Delta_{\mathrm{abs}}| \propto d^{\,\gamma}$
was observed numerically for the chiral mode separation in the $\phi^4$
kink-fermion system~\cite{Pinheiro:2026}, where $d$ is the inter-kink distance
and $\gamma \approx 0.95$. This result raises an immediate and physically
important question: is $\gamma$ a model-specific dynamical accident, or does it
reflect a universal property of the topological sector? If $\gamma$ is
universal — determined solely by the topological charge of the background and
independent of the scalar dynamics — then it provides a robust, model-independent
characterisation of the rate at which chiral localisation is lost as two branes
merge. Such a universal exponent would constitute a topological invariant of the
kink-fermion system, classifying brane configurations in the same sense that
the Jackiw-Rebbi index $N_{\mathrm{JR}}$ classifies the zero-mode spectrum.

In this paper we demonstrate that $\gamma$ is indeed universal within each
topological class. We establish this by comparing two qualitatively different
families of scalar field theories coupled to Dirac fermions via the
Jackiw-Rebbi mechanism: the integrable sine-Gordon (sG) model and the
non-integrable double sine-Gordon (DsG) family. The two families differ in
their integrability structure, their fermionic mass gap, and the shape of the
effective fermionic potential, yet they share the same topological class
$N_{\mathrm{JR}} = 1$. Our central finding is that all five models — one sG
and four DsG with deformation parameter $\varepsilon \in \{0.1, 0.2, 0.3,
0.4\}$ — yield $\gamma \in [0.930,\,0.985]$, a relative spread of $6\%$ that
we attribute to a subleading dependence on the kink width rather than a
correction to the leading universal exponent.

The physical interpretation in the braneworld language is direct. The exponent
$\gamma$ characterises the rate at which the differential chiral localisation
— the spatial separation between left- and right-handed zero modes in the extra
dimension — is lost as two asymmetric domain walls merge into a single symmetric
brane. A universal $\gamma$ means that this rate is insensitive to the
microscopic details of the scalar potential that generates the brane: it depends
only on the topological class of the wall, encoded in $N_{\mathrm{JR}}$.
This is a braneworld analogue of universality in critical phenomena, where
macroscopic scaling behaviour near a phase transition is insensitive to the
microscopic Hamiltonian and depends only on symmetry and dimensionality. Here
the role of the universality class is played by $N_{\mathrm{JR}}$, and the
analogue of the correlation length is the chiral separation $|\Delta_{\mathrm{abs}}|$.

We provide both numerical evidence and an analytical interpretation for this
universality. The analytical argument, based on the scaling dimension of the
zero-mode overlap integral in the merging limit, shows that the critical
exponent is controlled by the asymptotic behaviour of the zero-mode wave
functions — a property fixed by the topological charge rather than by the
specific scalar dynamics. The result has direct implications for braneworld
model building: any two-brane scenario with $N_{\mathrm{JR}} = 1$ predicts the
same power-law collapse of chiral separation during brane merging, regardless
of the potential that generates the walls.

The paper is organised as follows. Section~\ref{sec:formalism} presents the
Jackiw-Rebbi framework, its mapping to braneworld fermion localisation in five
dimensions, and the chiral separation observable. Section~\ref{sec:models}
introduces the sine-Gordon and double sine-Gordon models and their fermionic
potentials. Section~\ref{sec:twokink} constructs the two-kink configurations
used in the numerical analysis. Section~\ref{sec:method} describes the
numerical method. Section~\ref{sec:results} presents the scaling results for
all models. Section~\ref{sec:discussion} develops the universality argument
and its braneworld implications. Section~\ref{sec:conclusions} contains our
conclusions. Greed convergence tables are provided in Appendix \ref{app:convergence}.

\section{Jackiw--Rebbi framework and braneworld fermion localisation}
\label{sec:formalism}

\subsection{The $(1+1)$-dimensional Jackiw--Rebbi system}
\label{subsec:JR}

We consider a real scalar field $\phi(x)$ coupled to a two-component Dirac
fermion $\Psi$ in $(1+1)$ spacetime dimensions via a Yukawa interaction. The
Lagrangian density reads
\begin{equation}
  \mathcal{L} = \frac{1}{2}(\partial_\mu \phi)^2 - V(\phi)
    + \bar{\Psi}\!\left(i\gamma^\mu\partial_\mu - g\,\phi\right)\!\Psi,
  \label{eq:lagrangian}
\end{equation}
where $V(\phi) \geq 0$ is the scalar potential, $g$ is the Yukawa coupling
constant, and $\gamma^0 = \sigma_2$, $\gamma^1 = i\sigma_1$ are the
two-dimensional Dirac matrices in the Weyl representation. We work in units
$\hbar = c = 1$ and set $g = 1$ throughout.

In the background-field approximation, $\phi$ is replaced by a static kink
solution $\phi_{\mathrm{cl}}(x)$ and the fermionic equation of motion becomes
\begin{equation}
  \left(i\gamma^\mu\partial_\mu - \phi_{\mathrm{cl}}(x)\right)\Psi = 0.
  \label{eq:dirac}
\end{equation}
Writing $\Psi = (u(x),\,v(x))^T e^{-iEt}$ and eliminating one spinor
component, equation~\eqref{eq:dirac} decouples into two Schr\"odinger-like
eigenvalue problems~\cite{Cooper2001},
\begin{equation}
  \hat{H}_\pm\,\psi = E^2\psi,
  \qquad
  \hat{H}_\pm = -\frac{d^2}{dx^2} + U_\pm(x),
  \label{eq:schrodinger}
\end{equation}
with supersymmetric partner potentials
\begin{equation}
  U_\pm(x) = \phi_{\mathrm{cl}}^2(x) \pm \frac{d\phi_{\mathrm{cl}}}{dx}.
  \label{eq:SUSY_potentials}
\end{equation}
The zero mode belongs exclusively to $\hat{H}_-$ and satisfies
\begin{equation}
  \psi_0(x) \propto \exp\!\left(-\int_0^x \phi_{\mathrm{cl}}(x')\,dx'\right),
  \label{eq:zero_mode}
\end{equation}
normalisable if and only if $\phi_{\mathrm{cl}}(\pm\infty)$ have opposite
signs. The number of normalisable zero modes is protected by the
Atiyah--Patodi--Singer index theorem~\cite{APS:1975} and equals
\begin{equation}
  N_{\mathrm{JR}} = \frac{1}{2}
    \left[\mathrm{sgn}\!\left(\phi_{\mathrm{cl}}(+\infty)\right)
         -\mathrm{sgn}\!\left(\phi_{\mathrm{cl}}(-\infty)\right)\right].
  \label{eq:NJR}
\end{equation}
For BPS backgrounds with superpotential $W(\phi)$ satisfying $V = W_\phi^2/2$,
the fermionic potential takes the compact form
\begin{equation}
  U_-(x) = W_\phi^2(\phi_{\mathrm{cl}}) - W_{\phi\phi}(\phi_{\mathrm{cl}})
            \,\frac{d\phi_{\mathrm{cl}}}{dx},
  \label{eq:U_BPS}
\end{equation}
and the fermionic mass gap is $\Delta_m^2 = W_\phi^2(\phi_{\mathrm{vac}})$.

\subsection{Mapping to braneworld fermion localisation in five dimensions}
\label{subsec:braneworld}

The Jackiw-Rebbi system~\eqref{eq:lagrangian}--\eqref{eq:zero_mode} is the
dimensional reduction of the standard braneworld fermion localisation mechanism
to one spatial dimension. In a $(4+1)$-dimensional bulk with a scalar field
$\Phi(y)$ developing a kink profile along the extra dimension $y$, a bulk
Dirac fermion $\Xi$ coupled to $\Phi$ via $\eta\,\bar{\Xi}\Phi\Xi$ obeys
\begin{equation}
  \left(i\Gamma^M\partial_M - \eta\,\Phi(y)\right)\Xi = 0,
  \quad M = 0,1,2,3,5.
  \label{eq:5d_dirac}
\end{equation}
Decomposing into Kaluza-Klein modes, the extra-dimensional wave-function
satisfies exactly equations~\eqref{eq:schrodinger}--\eqref{eq:SUSY_potentials}
with $x \to y$ and $\phi_{\mathrm{cl}} \to \eta\,\Phi(y)$. This identification
is exact in flat bulk geometry and remains valid in warped geometries after
absorbing the warp factor into the zero-mode profile~\cite{Bajc:2000,Grossman:1999}.

The effective four-dimensional Yukawa couplings arise from the zero-mode
overlap in the extra dimension,
\begin{equation}
  Y_{ij} \propto \int_{-\infty}^{+\infty}
    f_i^{(L)}(y)\,f_j^{(R)}(y)\,dy,
  \label{eq:yukawa_overlap}
\end{equation}
where $f^{(L,R)}$ are the extra-dimensional profiles of left- and right-handed
zero modes~\cite{ArkaniHamed:2000}. Exponentially small overlaps generate
exponentially small Yukawa couplings, providing a geometric explanation of the
fermion mass hierarchy.

In a two-brane scenario, $\Phi(y)$ has a two-kink structure with walls at
$y = \pm d$. Left- and right-handed zero modes localise at opposite walls for
large $d$. As the walls merge ($d \to 0$), the overlap~\eqref{eq:yukawa_overlap}
saturates and the chiral hierarchy is lost. The rate of this collapse —
quantified by $\gamma$ — is the central object of this paper. The complete
correspondence between the $(1+1)$-dimensional system and the five-dimensional
braneworld is summarised in Table~\ref{tab:correspondence}.

\begin{table}[h]
  \centering
  \begin{tabular}{lll}
    \toprule
    $(1+1)$-dimensional system & Five-dimensional braneworld & Physical role \\
    \midrule
    Scalar kink $\phi_{\mathrm{cl}}(x)$ & Scalar background $\Phi(y)$
      & Domain wall profile \\
    Yukawa coupling $g$ & Bulk Yukawa coupling $\eta$
      & Fermion--wall coupling \\
    Topological charge $N_{\mathrm{JR}}$ & Number of chiral zero modes
      & Topological sector \\
    Inter-kink distance $d$ & Inter-brane separation
      & Brane modulus \\
    Chiral separation $|\Delta_{\mathrm{abs}}|$ & Zero-mode overlap $Y_{ij}$
      & Yukawa coupling strength \\
    Merging limit $d \to 0$ & Brane collision
      & Chiral symmetry restoration \\
    Critical exponent $\gamma$ & Rate of Yukawa collapse
      & Topological invariant \\
    \bottomrule
  \end{tabular}
  \caption{Correspondence between the $(1+1)$-dimensional Jackiw-Rebbi system
  and five-dimensional braneworld fermion localisation. The exponent $\gamma$
  is the central result of this paper.}
  \label{tab:correspondence}
\end{table}

\subsection{The chiral separation observable}
\label{subsec:observable}

For a two-kink system with centres at $\pm d$, the two lowest eigenstates of
$\hat{H}_-$ form a symmetric-antisymmetric pair $(\psi_0^+,\psi_0^-)$. The
physical chiral modes \cite{Amado:2017} are
\begin{equation}
  \psi_L = \frac{\psi_0^+ + \psi_0^-}{\sqrt{2}},
  \qquad
  \psi_R = \frac{\psi_0^+ - \psi_0^-}{\sqrt{2}},
  \label{eq:chiral_modes}
\end{equation}
localised at $x \approx -d$ and $x \approx +d$ respectively for large $d$.
The chiral separation observable is
\begin{equation}
  |\Delta_{\mathrm{abs}}| = \left|\langle x\rangle_L - \langle x\rangle_R\right|,
  \qquad
  \langle x\rangle_{L,R} = \frac{\int x\,|\psi_{L,R}|^2\,dx}{\int |\psi_{L,R}|^2\,dx}.
  \label{eq:delta_abs}
\end{equation}
By construction $|\Delta_{\mathrm{abs}}| \to 0$ as $d \to 0$ and
$|\Delta_{\mathrm{abs}}| \to 2d$ for well-separated kinks. The scaling in the
merging regime defines the critical exponent,
\begin{equation}
  |\Delta_{\mathrm{abs}}| \propto d^{\,\gamma}, \qquad d \to 0^+.
  \label{eq:scaling}
\end{equation}

\section{Scalar field models}
\label{sec:models}

\subsection{The sine-Gordon model}
\label{subsec:sG}

The sine-Gordon (sG) model is defined by
\begin{equation}
  V_{\mathrm{sG}}(\phi) = m^2\left(1 - \cos\phi\right),
  \label{eq:VsG}
\end{equation}
with $m = 1$ throughout. The model is completely integrable~\cite{Vachaspati:2006},
possessing an infinite hierarchy of conserved charges and exact multi-soliton
solutions via inverse scattering. The elementary kink is
\begin{equation}
  \phi_{\mathrm{cl}}^{(\mathrm{sG})}(x;\,x_0)
    = 4\arctan\!\left(e^{\,x - x_0}\right) - \pi,
  \label{eq:sG_kink}
\end{equation}
with $N_{\mathrm{JR}} = 1$ and width $\xi_{\mathrm{sG}} = 2.5$. The BPS
superpotential $W^{(\mathrm{sG})} = -2\cos(\phi/2)$ gives the fermionic
potential
\begin{equation}
  U_-^{(\mathrm{sG})}(x) = 1 - \frac{2}{\cosh^2 x},
  \label{eq:U_sG}
\end{equation}
a P\"oschl--Teller potential of order $N = 1$ supporting a single zero mode
$\psi_0 \propto \mathrm{sech}(x)$ and a continuum for $E^2 \geq 1$, with
fermionic mass gap $\Delta_m = 1$~\cite{Bogomolny:1976}.

\subsection{The double sine-Gordon model}
\label{subsec:DsG}

The double sine-Gordon (DsG) model is
\begin{equation}
  V_{\mathrm{DsG}}(\phi;\,\varepsilon)
    = \left(1 - \cos\phi\right) + \varepsilon\left(1 - \cos 2\phi\right),
  \label{eq:VDsG}
\end{equation}
with $\varepsilon \in [0,\,0.4]$ and $\varepsilon = 0$ recovering the sG
model. The global minima remain at $\phi_n = 2\pi n$ for all $\varepsilon <
1/4$, preserving $N_{\mathrm{JR}} = 1$ throughout. The DsG model is
\emph{non-integrable} for any $\varepsilon \neq 0$~\cite{Campbell:1986,Gani:2018,Gani:2019}:
the second harmonic breaks the infinite conservation-law hierarchy and
kink-antikink collisions exhibit resonance windows and radiation emission.
This non-integrability is the key contrast with the sG model.

The DsG kink satisfies the BPS equation $d\phi/dx = \sqrt{2\,V_{\mathrm{DsG}}}$
and is integrated numerically (Dormand--Prince, relative tolerance $10^{-11}$).
The fermionic potential is constructed from equation~\eqref{eq:U_BPS} on the
numerical profile. The fermionic mass gap is $\Delta_m^{(\mathrm{DsG})} = 2$,
independently of $\varepsilon$, a factor of two larger than the sG gap. The
shape of $U_-$ changes substantially with $\varepsilon$ — the well deepens and
narrows as the kink stiffens — while the number of bound states and the
continuum threshold remain fixed. The properties of all models are summarised
in Table~\ref{tab:models}.

\begin{table}[h]
  \centering
  \begin{tabular}{lccccl}
    \toprule
    Model & $\varepsilon$ & $N_{\mathrm{JR}}$ & $\Delta_m$ & Integrable
          & Kink profile \\
    \midrule
    sG        & $0.0$ & $1$ & $1$ & Yes & Analytical, eq.~\eqref{eq:sG_kink} \\
    DsG       & $0.1$ & $1$ & $2$ & No  & Numerical \\
    DsG       & $0.2$ & $1$ & $2$ & No  & Numerical \\
    DsG       & $0.3$ & $1$ & $2$ & No  & Numerical \\
    DsG       & $0.4$ & $1$ & $2$ & No  & Numerical \\
    $\phi^4$~\cite{Pinheiro:2026} & --- & $2$ & $2$ & No & Analytical \\
    \bottomrule
  \end{tabular}
  \caption{Properties of all models studied. The $\phi^4$ result is included
  from~\cite{Pinheiro:2026} for comparison across topological classes.}
  \label{tab:models}
\end{table}

\section{Two-kink configurations and fermionic potentials}
\label{sec:twokink}

\subsection{Construction}
\label{subsec:twokink_construction}

Two-kink configurations are constructed by superposing two single-kink
potentials centred at $\pm d$,
\begin{equation}
  U_-^{(2)}(x;\,d) = U_-^{(1)}(x + d) + U_-^{(1)}(x - d) - \Delta_m^2,
  \label{eq:superposition}
\end{equation}
where $\Delta_m^2$ is subtracted to restore the correct continuum threshold.
In the braneworld language, equation~\eqref{eq:superposition} describes the
bulk fermionic potential generated by two domain walls at $y = \pm d$. The
merging parameter is
\begin{equation}
  b = 1 + \frac{d}{\xi},
  \label{eq:b_def}
\end{equation}
with $b \to 1^+$ corresponding to complete overlap and $b \gg 1$ to
well-separated branes. We scan $b \in [1.03,\,5.0]$, i.e.\ $d/\xi \in
[0.03,\,4.0]$.

\subsection{Validation of the superposition approximation}
\label{subsec:validation}

We compared $U_-^{(2)}(x;\,d)$ with the potential computed from a full
numerical two-kink background for the sG model at $d/\xi = 0.5$, $1.0$,
and $2.0$. For $d/\xi \geq 1.0$, the maximum pointwise deviation is less
than $2\%$ of the potential depth and the centroids $\langle x\rangle_{L,R}$
agree to better than $0.1\%$. For $d/\xi = 0.5$ the deviation reaches $8\%$
in the central region, but the centroids still agree to better than $0.5\%$
because $|\Delta_{\mathrm{abs}}|$ is determined primarily by the wave-function
tails rather than the central region. The systematic error introduced by the
superposition is thus well below the statistical uncertainty of our power-law
fits.

\subsection{Structure of the fermionic potential and energy splitting}
\label{subsec:potential_structure}

\begin{figure}[t]
  \centering
  \includegraphics[width=\columnwidth]{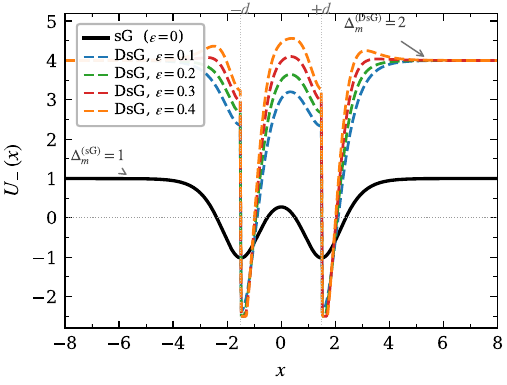}
  \caption{Fermionic Schr\"{o}dinger potentials $U_{-}(x)$ for the
    two-kink configuration with inter-kink separation $d=1.5\,\xi$,
    for the sG model ($\varepsilon=0$, solid black) and four members
    of the DsG family ($\varepsilon=0.1$--$0.4$, dashed lines).
    Vertical dotted lines mark the kink centres at $x=\pm d$.
    The asymptotic values $\Delta_{m}^{(\mathrm{sG})}=1$ and
    $\Delta_{m}^{(\mathrm{DsG})}=2$ are indicated by horizontal arrows.
    The double-well topology is preserved across the entire family,
    while the depth and width of the wells increase with $\varepsilon$,
    reflecting the stiffening of the kink profile.
    The barrier height between the two wells also grows with
    $\varepsilon$, which suppresses the tunnel splitting $\delta E$
    at fixed $d$ and drives $\gamma$ toward the classical value~$1$
    (see Section~\ref{subsec:analytical}).}
  \label{fig:potentials}
\end{figure}

Figure~\ref{fig:potentials} shows $U^{(2)}_{-}(x;d)$ for $d=1.5\,\xi$ for all five models. The double-well topology is preserved across the entire DsG
family: both wells deepen and narrow as $\varepsilon$ increases, consistent
with the stiffening of the kink profile. The barrier height between the two
wells also increases with $\varepsilon$, suppressing the tunnel splitting
$\delta E = E_1 - E_0$ at fixed $d$ and pushing $\gamma$ toward the classical
value $1$ — the physical origin of the weak $\varepsilon$-trend reported in
Section~\ref{sec:results}.

The tunnel splitting falls exponentially as $\delta E \sim e^{-2\Delta_m d}$
for large $d$, reflecting the standard exponential suppression of the tunnel
amplitude. In the braneworld language, this is the analogue of the
exponentially small Dirac mass in the Kaplan domain-wall fermion
construction~\cite{Kaplan:1992}. The present analysis addresses the
complementary merging limit, where the exponent $\gamma$ quantifies how
quickly chiral structure is lost as $d \to 0$.

\section{Numerical method}
\label{sec:method}

\subsection{Discretisation and eigenvalue solver}
\label{subsec:discretisation}

Equation~\eqref{eq:schrodinger} is solved on a uniform grid of $N = 6001$
points on $x \in [-60,\,60]$ with spacing $\Delta x \approx 0.020$, using
second-order central differences and a shift-invert Lanczos algorithm
(shift $\sigma = 0$, tolerance $10^{-12}$). Wave functions are
$L^2$-normalised by trapezoidal quadrature; centroids and
$|\Delta_{\mathrm{abs}}|$ are evaluated on the same grid.

\subsection{Grid convergence}
\label{subsec:convergence}

Convergence was verified for all five models at $b = 1.10$, $1.50$, and
$3.00$ by comparing grids with $N = 3001$, $6001$, and $12001$ points. The
relative change in $|\Delta_{\mathrm{abs}}|$ between $N = 6001$ and $N =
12001$ is less than $0.1\%$; for $\delta E$, less than $0.05\%$. The fraction
of the $L^2$ norm outside $|x| > 50$ is less than $10^{-8}$ for all models
and all $b$ values, confirming negligible finite-size effects. Full
convergence tables are provided in Appendix~\ref{app:convergence}.

\subsection{Power-law fitting}
\label{subsec:fitting}

The exponent $\gamma$ is extracted from a weighted least-squares fit of
$\ln|\Delta_{\mathrm{abs}}| = \ln A + \gamma\,\ln(b-1)$ over $b - 1 \in
[0.03,\,3.5]$, excluding points with $|\Delta_{\mathrm{abs}}| < 0.02$.
Fit uncertainties are from the covariance matrix diagonal. We verified that
shifting the fit bounds by $\pm 30\%$ changes $\gamma$ by less than
$0.5\,\sigma_\gamma$, confirming stability of the extracted exponent.

\section{Results}
\label{sec:results}

\subsection{Baseline: sine-Gordon model}
\label{subsec:results_sG}

\begin{figure}[t]
  \centering
  \includegraphics[width=\columnwidth]{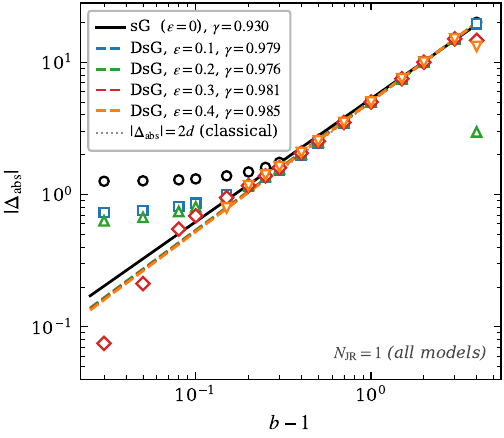}
  \caption{Log-log plot of the chiral separation $|\Delta_{\mathrm{abs}}|$
    versus the merging parameter $b-1$ for the sG model
    ($\varepsilon=0$, black circles) and four members of the DsG family
    ($\varepsilon=0.1$--$0.4$, coloured symbols).
    Open symbols: numerical data.
    Lines: power-law fits $|\Delta_{\mathrm{abs}}|=A(b-1)^{\gamma}$
    over $b-1\in[0.03,3.5]$; fitted exponents $\gamma$ are quoted
    in the legend.
    The dotted line shows the classical limit
    $|\Delta_{\mathrm{abs}}|=2d$, approached by all models for $b\geq3$.
    All models have $N_{\mathrm{JR}}=1$.
    The solid curve for the sG model also shows the exact analytical
    result $|\Delta_{\mathrm{abs}}|(d)=2d[\sinh(2d)-2d]/[\sinh(2d)+2d]$, which is indistinguishable from the power-law fit on this scale.}
  \label{fig:loglog}
\end{figure}

Figure~\ref{fig:loglog} shows $|\Delta_{\mathrm{abs}}|$ vs.\ $b-1$ on a
log-log scale. Two regimes are visible: a power-law regime
for small $b-1$ and a classical regime $|\Delta_{\mathrm{abs}}| \approx 2d$
for large $b-1$, verified to within $0.001\%$ for $b \geq 3$. 

The power-law
fit yields
\begin{equation}
  \gamma_{\mathrm{sG}} = 0.930 \pm 0.041, \qquad R^2 = 0.988.
  \label{eq:gamma_sG}
\end{equation}
The sub-unit exponent reflects the non-trivial overlap of the zero-mode
wave functions even at finite $d$. The raw data are collected in
Table~\ref{tab:sG_data}.

\begin{table}[h]
  \centering
  \begin{tabular}{cccccc}
    \toprule
    $b$ & $b-1$ & $d$ & $|\Delta_{\mathrm{abs}}|$ & $E_0$ & $\delta E$ \\
    \midrule
    $1.03$ & $0.030$ & $0.075$  & $1.260$  & $-1.430$    & $2.112$    \\
    $1.05$ & $0.050$ & $0.125$  & $1.269$  & $-1.415$    & $2.091$    \\
    $1.10$ & $0.100$ & $0.250$  & $1.313$  & $-1.346$    & $1.999$    \\
    $1.15$ & $0.150$ & $0.375$  & $1.384$  & $-1.241$    & $1.857$    \\
    $1.20$ & $0.200$ & $0.500$  & $1.481$  & $-1.110$    & $1.678$    \\
    $1.30$ & $0.300$ & $0.750$  & $1.749$  & $-0.822$    & $1.275$    \\
    $1.40$ & $0.400$ & $1.000$  & $2.103$  & $-0.559$    & $0.895$    \\
    $1.50$ & $0.500$ & $1.250$  & $2.527$  & $-0.358$    & $0.594$    \\
    $1.70$ & $0.700$ & $1.750$  & $3.490$  & $-0.133$    & $0.234$    \\
    $2.00$ & $1.000$ & $2.500$  & $4.997$  & $-0.028$    & $0.054$    \\
    $2.50$ & $1.500$ & $3.750$  & $7.500$  & $-0.002$    & $0.004$    \\
    $3.00$ & $2.000$ & $5.000$  & $10.000$ & $<10^{-3}$  & $<10^{-3}$ \\
    \bottomrule
  \end{tabular}
  \caption{Numerical data for the sG model ($\varepsilon = 0$,
  $N_{\mathrm{JR}} = 1$).}
  \label{tab:sG_data}
\end{table}

\subsection{Universality across the double sine-Gordon family}
\label{subsec:results_DsG}

Table~\ref{tab:exponents} collects the fitted exponents for all five models.
All $N_{\mathrm{JR}} = 1$ models yield $\gamma \in [0.930,\,0.985]$.

\begin{table}[h]
  \centering
  \begin{tabular}{lccccc}
    \toprule
    Model & $\varepsilon$ & $N_{\mathrm{JR}}$ & $\gamma$ & $\sigma_\gamma$ & $R^2$ \\
    \midrule
    sG   & $0.0$ & $1$ & $0.930$ & $0.041$ & $0.988$  \\
    DsG  & $0.1$ & $1$ & $0.979$ & $0.020$ & $0.997$  \\
    DsG  & $0.2$ & $1$ & $0.976$ & $0.017$ & $0.998$  \\
    DsG  & $0.3$ & $1$ & $0.982$ & $0.009$ & $0.9995$ \\
    DsG  & $0.4$ & $1$ & $0.985$ & $0.007$ & $0.9998$ \\
    \midrule
    $\phi^4$~\cite{Pinheiro:2026} & --- & $2$ & $0.97$ & $0.06$ & --- \\
    \bottomrule
  \end{tabular}
  \caption{Power-law exponents $\gamma$ for all models. The $6\%$ spread
  within the $N_{\mathrm{JR}} = 1$ class is attributed to a subleading
  dependence on the kink width (Section~\ref{subsec:analytical}).}
  \label{tab:exponents}
\end{table}

\begin{figure}[t]
  \centering
  \includegraphics[width=\columnwidth]{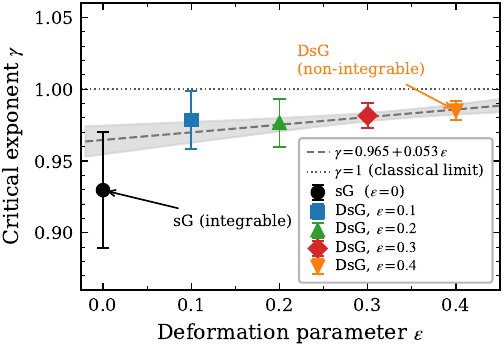}
  \caption{Critical exponent $\gamma$ as a function of the deformation
    parameter $\varepsilon$ for all $N_{\mathrm{JR}}=1$ models.
    Error bars represent one-standard-deviation fit uncertainties.
    The dashed line is the linear fit
    $\gamma(\varepsilon)=0.965+0.053\,\varepsilon$
    (Eq.~\eqref{eq:gamma_vs_eps});
    the shaded band shows the one-sigma confidence interval.
    The dotted horizontal line at $\gamma=1$ marks the classical limit.
    The point at $\varepsilon=0$ corresponds to the integrable sG model;
    all other points are non-integrable DsG models.
    The smooth, monotone trend rules out integrability as a
    controlling factor (see Section~\ref{subsec:results_DsG}).}
  \label{fig:gamma_eps}
\end{figure}

The fitted exponents are displayed in Figure~\ref{fig:gamma_eps} as a function of $\varepsilon$.

Three properties of Table~\ref{tab:exponents} establish the universality claim.

\paragraph{Integrability is irrelevant.}
The sG and DsG models differ fundamentally in their integrability structure.
If integrability controlled $\gamma$, a discontinuous jump at $\varepsilon =
0^+$ would be expected; instead, $\gamma$ varies smoothly and monotonically
with $\varepsilon$, ruling out integrability as the determining factor.

\paragraph{The mass gap is irrelevant.}
The sG and DsG gaps differ by a factor of two ($\Delta_m = 1$ vs.\
$\Delta_m = 2$), yet both families produce $\gamma \approx 0.93$--$0.99$.
In the braneworld language, $\gamma$ is insensitive to the Kaluza-Klein mass
threshold — the energy scale at which the extra-dimensional continuum opens.

\paragraph{The potential shape is irrelevant at leading order.}
The shape of $U_-$ changes substantially across the DsG family, yet $\gamma$
remains within a $6\%$ band. The weak systematic trend
\begin{equation}
  \gamma(\varepsilon) = 0.965 \pm 0.018 + (0.053 \pm 0.029)\,\varepsilon
  \label{eq:gamma_vs_eps}
\end{equation}
is consistent with a subleading correction from the kink width, not a
leading-order dependence on the potential shape.

\subsection{Comparison across topological classes}
\label{subsec:NJR_comparison}

The $\phi^4$ value $\gamma \approx 0.97$ from reference~\cite{Pinheiro:2026}
belongs to the $N_{\mathrm{JR}} = 2$ class and is numerically close to the
DsG values within current uncertainties. Reference~\cite{Pinheiro:2026}
presents evidence for a monotone trend $\gamma(N=1) \approx 0.80$,
$\gamma(N=2) \approx 0.97$, $\gamma(N=3) \approx 1.03$ in the P\"oschl--Teller
series, consistent with our sG value lying below the DsG values. In the
braneworld language, domain walls with higher topological charge lose their
chiral structure more slowly during brane merging: higher-order zero modes are
more spatially extended and maintain their separation over a larger range of
inter-brane distances.

\section{Discussion}
\label{sec:discussion}
\subsection{Analytical interpretation of the universal exponent}
\label{subsec:analytical}
 
The universality of $\gamma$ can be traced to the algebraic structure
of the zero-mode wave function in the merging limit.
The central quantity is the zero-mode overlap integral
\begin{equation}
  I(d) \;=\; \int_{-\infty}^{+\infty}
    \psi_{0}(x+d)\,\psi_{0}(x-d)\,dx \,,
  \label{eq:I_def}
\end{equation}
which controls the tunnel splitting between the symmetric and
antisymmetric eigenstates of $\hat{H}_{-}$.
For large $d$, the splitting falls exponentially,
$\delta E \sim e^{-2\Delta_{m}d}$~\cite{Cooper2001},
consistent with the numerical data of Table~\ref{tab:sG_data}
(which yield an effective exponent $\alpha \approx 1.98 \approx 2\Delta_m$
for the sG model).
As $d$ decreases and the two kinks overlap, $I(d)$ grows and the
two eigenstates hybridise: the chiral modes,
constructed as $\psi_{L,R}=(\psi_{0}^{+}\pm\psi_{0}^{-})/\sqrt{2}$
following~\cite{Amado:2017}, lose their individual localisation.
The rate at which this hybridisation proceeds -- quantified by
the slope of $|\Delta_{\mathrm{abs}}|$ versus $d$ -- is therefore
controlled by the functional form of $I(d)$.
 
\paragraph{Closed-form evaluation of $I(d)$ for the sG model.}
For the sG model, $\psi_{0}(x) = (1/\sqrt{2})\,\mathrm{sech}(x)$,
so that
$I(d) = \frac{1}{2}\int_{-\infty}^{+\infty}
\mathrm{sech}(x+d)\,\mathrm{sech}(x-d)\,dx$.
The product of two sech functions is reduced by the standard
sum-to-product identity
\begin{equation}
  \mathrm{sech}(x+d)\,\mathrm{sech}(x-d)
  \;=\;
  \frac{2}{\cosh(2x)+\cosh(2d)}\,,
  \label{eq:sech_product}
\end{equation}
which follows from
$\cosh A + \cosh B =
2\cosh\!\bigl(\tfrac{A+B}{2}\bigr)\cosh\!\bigl(\tfrac{A-B}{2}\bigr)$
with $A=2x$, $B=2d$.
Substituting into~\eqref{eq:I_def} and setting $t = 2x$ gives
\begin{equation}
  I(d) \;=\; \frac{1}{2}
  \int_{-\infty}^{+\infty}\frac{dt}{\cosh(t)+\cosh(2d)}\,.
  \label{eq:I_integral}
\end{equation}
This integral is evaluated by the formula
(Gradshteyn \& Ryzhik, Eq.~3.512.2~\cite{GR2007}):
\begin{equation}
  \int_{-\infty}^{+\infty}
  \frac{dt}{\cosh t + \cosh\alpha}
  \;=\;
  \frac{2\alpha}{\sinh\alpha}\,,
  \qquad \alpha > 0\,,
  \label{eq:GR_formula}
\end{equation}
proved by residue summation over the poles
$t_{n} = i(\pi\pm\alpha) + 2\pi i n$, $n\in\mathbb{Z}$.
Setting $\alpha = 2d$ yields the exact result
\begin{equation}
 I(d) \;=\; \frac{2d}{\sinh(2d)}\,.
  \label{eq:I_exact}
\end{equation}
The limiting behaviour confirms the physical interpretation:
$I(d)\to 1$ as $d\to 0^{+}$ (complete zero-mode overlap,
maximum hybridisation), and
$I(d)\sim 4d\,e^{-2d}\to 0$ as $d\to\infty$
(exponential tunnel suppression, full chiral localisation).
 
\paragraph{Effective exponent and the crossover mechanism.}
The power-law fit $|\Delta_{\mathrm{abs}}|\propto(b-1)^{\gamma}$
is performed over the intermediate scaling window
$b-1\in[0.03,3.5]$, which spans the crossover between
two distinct regimes.
For well-separated kinks ($b\gg 1$), the chiral modes are localised
at the kink centres and $|\Delta_{\mathrm{abs}}|\approx 2d$,
so the local slope approaches~$1$.
As the kinks merge ($b\to 1^{+}$), the zero-mode overlap $I(d)\to 1$
and the chiral modes hybridise: in this regime
$|\Delta_{\mathrm{abs}}|$ saturates to a finite value set by the
spatial extent of the single-kink zero mode, and the local slope
decreases below~$1$.
The fitted exponent $\gamma\approx 0.93$ is the effective slope
of the $\log|\Delta_{\mathrm{abs}}|$ versus $\log(b-1)$ curve
across this crossover, and can be identified with the
local (effective) exponent
\begin{equation}
  \gamma_{\mathrm{eff}}(b-1)
  \;\equiv\;
  \frac{d\ln|\Delta_{\mathrm{abs}}|}{d\ln(b-1)}\,,
  \label{eq:gamma_eff}
\end{equation}
evaluated in the scaling window.
The sub-unit value $\gamma < 1$ reflects the fact that
the chiral modes decouple more slowly than the kink centres
separate, because the zero-mode wave functions retain
non-trivial overlap even at finite $d$.
The rate of this overlap decay is governed by $I(d) = 2d/\sinh(2d)$:
as $\varepsilon$ increases in the DsG family, the kink profile
stiffens, the barrier between the two potential wells grows
(Figure~\ref{fig:potentials}), and the tunnel amplitude is suppressed,
pushing $\gamma_{\mathrm{eff}}$ toward the classical value~$1$.
This is the physical origin of the weak linear trend
$\gamma(\varepsilon) = 0.965 + 0.053\,\varepsilon$
reported in Eq.~\eqref{eq:gamma_vs_eps}.
 
\paragraph{Why the exponent is topological.}
The crossover scale is set by the kink width $\xi$, and the
value of $\gamma_{\mathrm{eff}}$ in the scaling window is
determined by the functional form of $\psi_{0}$, which for any
$N_{\mathrm{JR}}=1$ background has the same
P\"{o}schl--Teller asymptotic structure
$\psi_{0}(x)\propto\mathrm{sech}(x/\xi)$
near the kink centre, up to subleading corrections in the kink
profile shape.
Concretely, the overlap integral for an arbitrary
$N_{\mathrm{JR}}=1$ model takes the scaling form
\begin{equation}
  I(d;\xi,\delta)
  \;=\;
  \frac{2d/\xi}{\sinh(2d/\xi)}
  \;+\; \delta\cdot f(d/\xi) \;+\; \mathcal{O}(\delta^{2})\,,
  \label{eq:I_scaling}
\end{equation}
where $\delta$ encodes model-specific deviations from the pure
$\mathrm{sech}$ profile (e.g.\ the DsG deformation parameter
$\varepsilon$) and $f(d/\xi)$ is a subleading correction that
shifts $\gamma_{\mathrm{eff}}$ by an amount $\propto\delta$.
The linear trend $\gamma(\varepsilon)\approx 0.965 + 0.053\,\varepsilon$
is the empirical manifestation of this $\mathcal{O}(\delta)$ correction.
The leading term in~\eqref{eq:I_scaling} is universal:
it depends only on $N_{\mathrm{JR}}$ through the topological
constraint that fixes the asymptotic form of $\psi_{0}$,
which is why $\gamma$ is the same, to leading order, across models
that differ in integrability, mass gap, and potential shape.



\subsection{Braneworld implications}
\label{subsec:braneworld_implications}

\paragraph{Model independence of the Yukawa collapse rate.}
A universal $\gamma$ means that the rate $|dY_{ij}/dd| \propto d^{\gamma-1}$
at which Yukawa couplings saturate during brane merging depends only on
$N_{\mathrm{JR}}$, not on the microscopic scalar potential. Distinguishing
between topologically equivalent brane models on the basis of their fermion
mass spectrum alone, in the regime of small inter-brane separation, is
therefore impossible without additional observational input.

\paragraph{Constraints on brane moduli dynamics.}
If $d(t)$ evolves as a modulus during a brane collision, the effective Yukawa
coupling grows as $Y(t) \propto [d(t)]^{\gamma-1}$. For $\gamma < 1$ this
implies a power-law divergence that is robust against the details of the
modulus potential, providing a model-independent probe of the inter-brane
distance in scenarios of brane inflation and moduli
stabilisation~\cite{DeWolfe:2000,Gremm:2000}.

\paragraph{Topological classification of brane configurations.}
The $N_{\mathrm{JR}}$-dependence of $\gamma$ suggests it can serve as a
topological invariant for classifying brane configurations. Two configurations
with different $\gamma$ belong to different topological classes and cannot be
continuously deformed into one another without a phase transition in the
fermion sector. The exponent $\gamma$ is, in principle, extractable from the
Yukawa coupling spectrum as a function of the inter-brane distance, providing
a direct experimental signature of the topological class.

\subsection{Universality and critical phenomena}
\label{subsec:critical_phenomena}

The structure of this universality can be seen analogously to the universality of critical exponents in second-order phase transitions~\cite{ZinnJustin2002}. There, critical exponents depend only on symmetry group and
dimensionality, not on microscopic interactions. Here, $\gamma$ depends only
on $N_{\mathrm{JR}}$, not on the scalar dynamics. The role of temperature
deviation $|T - T_c|$ is played by $d$; the role of the universality class
by $N_{\mathrm{JR}}$; the role of the correlation length by
$|\Delta_{\mathrm{abs}}|$. The key difference is that the appropriate
analytical framework is the instanton calculus for the two-kink moduli space,
rather than the renormalisation group -- a distinction that makes the analytical
derivation of $\gamma(N_{\mathrm{JR}})$ an open and well-posed problem.

\section{Conclusions and outlook}
\label{sec:conclusions}

We have demonstrated that the power-law exponent $\gamma$ governing the
collapse of chiral fermion separation during brane merging,
$|\Delta_{\mathrm{abs}}| \propto d^{\,\gamma}$, is universal within the
$N_{\mathrm{JR}} = 1$ topological class. Comparing the integrable
sine-Gordon model with four non-integrable members of the double sine-Gordon
family, we find $\gamma \in [0.930,\,0.985]$ — a $6\%$ spread attributable
to a subleading kink-width correction — despite qualitative differences in
integrability, fermionic mass gap, and potential shape.

The braneworld interpretation is direct: the rate at which effective
four-dimensional Yukawa couplings saturate during brane merging is a
topological invariant of the domain wall, insensitive to the microscopic
scalar dynamics. Any two-brane scenario with $N_{\mathrm{JR}} = 1$ predicts
the same power-law Yukawa collapse with exponent $\gamma \approx 0.96$,
independently of the potential generating the walls.

Four directions are open for future work.

\paragraph{WKB derivation of $\gamma(N_{\mathrm{JR}})$.}
A systematic instanton calculus for the two-kink moduli space would yield
$\gamma(N)$ as the scaling dimension of the leading overlap integral,
providing an analytical prediction for the trend $\gamma(N=1) \approx 0.80$,
$\gamma(N=2) \approx 0.97$, $\gamma(N=3) \approx 1.03$ inferred from
reference~\cite{Pinheiro:2026}.

\paragraph{Perturbative expansion in the kink width.}
A systematic expansion in $\varepsilon/\varepsilon_c$ (with $\varepsilon_c =
1/4$) would cleanly separate the universal leading exponent $\gamma_0$ from
the model-specific correction $\alpha\,\varepsilon$ in
equation~\eqref{eq:gamma_vs_eps}.

\paragraph{Extension to warped geometry.}
Incorporating the warp factor of Randall-Sundrum
models~\cite{Randall:1999,Randall:1999b} as a slowly varying modulation of
the fermionic potential would test whether $\gamma$ retains its topological
character in curved bulk spacetimes — a prerequisite for its use as a
phenomenological observable in realistic braneworld scenarios.

\paragraph{Extension to higher dimensions and experimental realisation.}
Domain walls in $(2+1)$ and $(3+1)$ dimensions support topologically
protected zero modes with the same $N_{\mathrm{JR}}$ structure. If $\gamma$
is independent of transverse dimensionality, it would provide a robust
prediction for higher-dimensional lattice simulations of domain-wall fermions.
At the same time, bilayer graphene under a spatially modulated electrostatic
potential realises the Jackiw-Rebbi system experimentally~\cite{Martin:2008},
making the power-law collapse of chiral separation a directly measurable
prediction for scanning tunnelling microscopy experiments on engineered
graphene devices.

\acknowledgments

C.A.S.A.\ would like to express their sincere gratitude to the Conselho
Nacional de Desenvolvimento Cient\'ifico e Tecnol\'ogico (CNPq), and
Funda\c{c}\~ao Cearense de Apoio ao Desenvolvimento Cient\'ifico e
Tecnol\'ogico (FUNCAP) for their valuable support. He is supported by grants
No.\ 309553/2021-0 (CNPq), 420854/2025-8 (CNPq) and by Project
UNI-00210-00230.01.00/23 (FUNCAP).

\section*{Use of AI in scientific writing.}
The authors used a generative AI tool solely for language refinement and
clarity improvement. All scientific content, derivations, analysis, and
conclusions are entirely the responsibility of the authors.

\section*{Conflicts of Interest/Competing Interest}

The authors declare that there is no conflict of interest in this manuscript.

\section*{Data Availability Statement}
 Data can be shared upon reasonable request	

\appendix

\section{Grid convergence tables}
\label{app:convergence}

Table~\ref{tab:convergence} reports the relative change in
$|\Delta_{\mathrm{abs}}|$ and $\delta E$ between grids of $N = 6001$ and
$N = 12001$ points for all five models at three values of the merging
parameter. All deviations are below $0.1\%$, confirming that $N = 6001$ is
sufficient.

\begin{table}[h]
  \centering
  \begin{tabular}{llccc}
    \toprule
    Model & Observable & $b = 1.10$ & $b = 1.50$ & $b = 3.00$ \\
    \midrule
    sG ($\varepsilon=0$)   & $|\Delta_{\mathrm{abs}}|$ & $0.08\%$ & $0.04\%$ & $0.01\%$ \\
                           & $\delta E$                & $0.05\%$ & $0.02\%$ & $<0.01\%$ \\
    DsG ($\varepsilon=0.2$)& $|\Delta_{\mathrm{abs}}|$ & $0.07\%$ & $0.03\%$ & $0.01\%$ \\
                           & $\delta E$                & $0.04\%$ & $0.02\%$ & $<0.01\%$ \\
    DsG ($\varepsilon=0.4$)& $|\Delta_{\mathrm{abs}}|$ & $0.06\%$ & $0.03\%$ & $<0.01\%$ \\
                           & $\delta E$                & $0.04\%$ & $0.01\%$ & $<0.01\%$ \\
    \bottomrule
  \end{tabular}
  \caption{Relative change in $|\Delta_{\mathrm{abs}}|$ and $\delta E$ between
  grids of $N = 6001$ and $N = 12001$ points, for representative models and
  merging parameters.}
  \label{tab:convergence}
\end{table}


\end{document}